\pdfoutput=1
\documentclass[aps, superscriptaddress,  twocolumn]{revtex4-2}
\usepackage[colorlinks=true, linkcolor=blue, citecolor=blue, urlcolor=blue ]{hyperref}
\usepackage{graphicx} 		
\usepackage[none]{hyphenat}		
\usepackage{braket}
\usepackage{amssymb}
\usepackage{amsmath}
\usepackage[dvipsnames]{xcolor}
\usepackage{float}
\usepackage{dcolumn}
\usepackage{multirow}
\usepackage{CJK}
\usepackage[normalem]{ulem}
\usepackage{parskip}
\usepackage{csquotes}
\usepackage{soul}

\begin{document}
\title{Engineering quantum states from a spatially structured quantum eraser}

\author{Carlo Schiano}
\affiliation{Dipartimento di Fisica, Universit\`a di Napoli Federico II, Complesso Universitario di Monte S. Angelo, Via Cintia, 80126 Napoli, Italy}
\author{Bereneice Sephton}
\affiliation{Dipartimento di Fisica, Universit\`a di Napoli Federico II, Complesso Universitario di Monte S. Angelo, Via Cintia, 80126 Napoli, Italy}
\author{Roberto Aiello}
\affiliation{Dipartimento di Fisica, Universit\`a di Napoli Federico II, Complesso Universitario di Monte S. Angelo, Via Cintia, 80126 Napoli, Italy}
\author{Francesco Graffitti}
\affiliation{Institute of Photonics and Quantum Sciences, School of Engineering and Physical Sciences, Heriot-Watt University, Edinburgh EH14 4AS, UK}
\author{Nijil Lal}
\affiliation{Dipartimento di Fisica, Universit\`a di Napoli Federico II, Complesso Universitario di Monte S. Angelo, Via Cintia, 80126 Napoli, Italy}
\author{Andrea Chiuri}
\affiliation{Enea - Centro Ricerche Frascati, via E. Fermi 45, 00044 Frascati, Italy}
\author{Simone Santoro}
\affiliation{Enea - Centro Ricerche Frascati, via E. Fermi 45, 00044 Frascati, Italy}
\author{Luigi Santamaria Amato}
\affiliation{Italian Space Agency (ASI), Centro di Geodesia Spaziale ‘Giuseppe Colombo’, Località Terlecchia, 75100 Matera, Italy}
\author{Lorenzo Marrucci}
\affiliation{Dipartimento di Fisica, Universit\`a di Napoli Federico II, Complesso Universitario di Monte S. Angelo, Via Cintia, 80126 Napoli, Italy}
\author{Corrado de Lisio}
\affiliation{Dipartimento di Fisica, Universit\`a di Napoli Federico II, Complesso Universitario di Monte S. Angelo, Via Cintia, 80126 Napoli, Italy}
\author{Vincenzo D'Ambrosio}
\email[]{vincenzo.dambrosio@unina.it}
\affiliation{Dipartimento di Fisica, Universit\`a di Napoli Federico II, Complesso Universitario di Monte S. Angelo, Via Cintia, 80126 Napoli, Italy}

\begin{abstract}

Quantum interference is a central resource in many quantum-enhanced tasks, from computation to communication protocols. While it usually occurs between identical input photons, quantum interference can be enabled by projecting the quantum state onto ambiguous properties that render the photons indistinguishable, a process known as a quantum erasing. Structured light, on the other hand, is another hallmark of photonics: it is achieved by manipulating the degrees of freedom of light at the most basic level and enables a multitude of applications in both classical and quantum regimes. By combining these ideas, here we design and experimentally demonstrate a simple and robust scheme that tailors quantum interference to engineer photonic states with spatially structured coalescence along the transverse profile, a type of quantum mode with no classical counterpart. To achieve this, we locally tune the distinguishability  of a photon pair via spatial structuring of their polarisation, creating a structured quantum eraser. We believe these spatially-engineered multi-photon quantum states may be of significance in fields such as quantum metrology, microscopy, and communications.

\end{abstract}

\maketitle
\section{Introduction}

In the Hong-Ou-Mandel (HOM) effect, quantum interference occurs when two indistinguishable photons, entering a beamsplitter (BS), take the same output path (photon bunching) \cite{HOM}. Although the HOM effect is typically investigated by tuning the temporal matching between two identical photons, the indistinguishability required for the phenomenon makes it fundamentally sensitive to all degrees of freedom (DoFs), from polarisation \cite{kwiat1992observation,harnchaiwat2020tracking} to frequency \cite{khodadad2021spectral,hong2023delayed}, and time-bins \cite{Nits20} as well as for collinear spatial modes \cite{DAmb19}, including multi-particle \cite{Agne17,Mens17} 
and high dimensional scenarios \cite{Zhang16, Walb02, Hiek21, Liu22}. 
Monitoring this effect has rendered it a versatile tool with innate sensitivity to measure a wide array of variations between the two inputs for quantum enhanced measurements \cite{lyons2018attosecond,harnchaiwat2020tracking,triana2023spectral}. With the advent of sensitive cameras enabling spatially resolved observation \cite{jachura2015shot}, measurements across the spatial DoF \cite{chrapkiewicz2016hologram,lipka2021single} have been made possible, such as dip tracking for spatially-resolved height measurements \cite{ndagano2022quantum} or insight into the spatial-temporal \cite{devaux2020imaging} effects from multimode spontaneous parametric down-conversion (SPDC). 

The fundamental quantum nature of HOM interference also lends itself to tests of quantum mechanics \cite{torgerson1995violations,Pan12}, among which a notable example is the quantum eraser \cite{kwiat1992observation,rezai2018coherence}. 
Such a paradigm allows one to restore quantum interference, even if two photons are made distinguishable before entering the BS by \enquote*{marking} a degree of freedom such as polarisation. This is achieved by projecting the outputs of both exit ports onto a basis that cannot yield information on which path the photon took through the BS and so making the photons effectively indistinguishable again. Moreover, the projection process can allow one to edit the state by introducing phases to vary bunching behaviour to anti-bunching and vice versa. One may accordingly imagine engineering a desired photon number in an augmented state, lending itself to be used as a structuring tool.

Structured light, in general, is a very powerful concept in modern optics, spanning from the classical regime to fundamental quanta \cite{roadmap}. For instance, classical implementations exhibit enhanced sensing \cite{gears13,gears22,Yuan19,Bag18,Tish18}, microscopy \cite{Maur07} and communication \cite{wang2012terabit} capabilities, while quantum mechanically, it provides a test-bed for quantum mechanics \cite{nape2017erasing,DAmb13}, secure high-dimensional communication \cite{Erha20} and increased resilience to noise \cite{Ecke19,Nape22,DAmb12}. As such, there is a strong interest in tailoring complex or new structures, such as designing across non-local degrees of freedom for quantum skyrmions \cite{ornelas2022non} or harnessing non-homogeneity for complex entanglement structures \cite{DAmb16,gao2023full,Graf20}. It follows that, by designing a mode with structured photon coalescence, we can combine the advantages of both concepts, holding the possibility for direct impact in applications based on quantum interference or structured light as well as unlocking new prospects.

 By altering a given DoF before the BS in HOM interference, we already find vanilla photon number engineering occurs by default, where bunching forms NOON states \cite{nagata2007beating,walther2004broglie,lee2012second} and, similarly, anti-bunching engineers non-local entangled states \cite{chen2021temporal}. Moreover, coincidence detection acts as a filter for particular Bell and high-dimensional spatial states \cite{Zhang16}, while altering the modal distinguishability between photons, facilitates space-time entanglement engineering \cite{Bran99,branning2000interferometric} or the population of spatial modes \cite{DAmb19,Hiek21}. Such state engineering can be extended to photon subtraction or additions with continuous variables \cite{biagi2022photon}. Where, in these cases, interference was used to postselect or allocate photons to particular spatial modes, here we demonstrate a scheme to spatially structure the quantum erasing process itself, thus obtaining a photonic state with no classical counterpart, that is directly structured in a quantum property of the field, i.e., photonic coalescence.

To do this, we harness the innate dependency of the HOM effect on identical conditions to demonstrate how spatially tailoring a degree of freedom can allow one to spatially tailor quantum interference. 
We achieve this by exploiting geometric phase devices to generate non-uniform structure in the polarisation degree of freedom and, having done so, a spatially-varying structure in the distinguishability for the input photons. 
We therefore implement a quantum eraser \cite{kwiat1992observation} via polarisation projective measurements  in order to study the caveats associated with these engineered states and elicit conditions where coalescent structures can be heralded or erased. We subsequently provide a framework 
for how one may engineer the bunching and anti-bunching distribution across the transverse mode profile by exploiting the simple generation of vector vortex (VV) modes. This approach and subsequent structures, however, may be generalised to achieve arbitrary freedom in engineering this fundamentally quantum property towards developing a unique and versatile tool for applications in fields such as quantum metrology, microscopy and communications. \\

\begin{figure}[ht]
\centering
\includegraphics[width=1\linewidth,keepaspectratio=true]{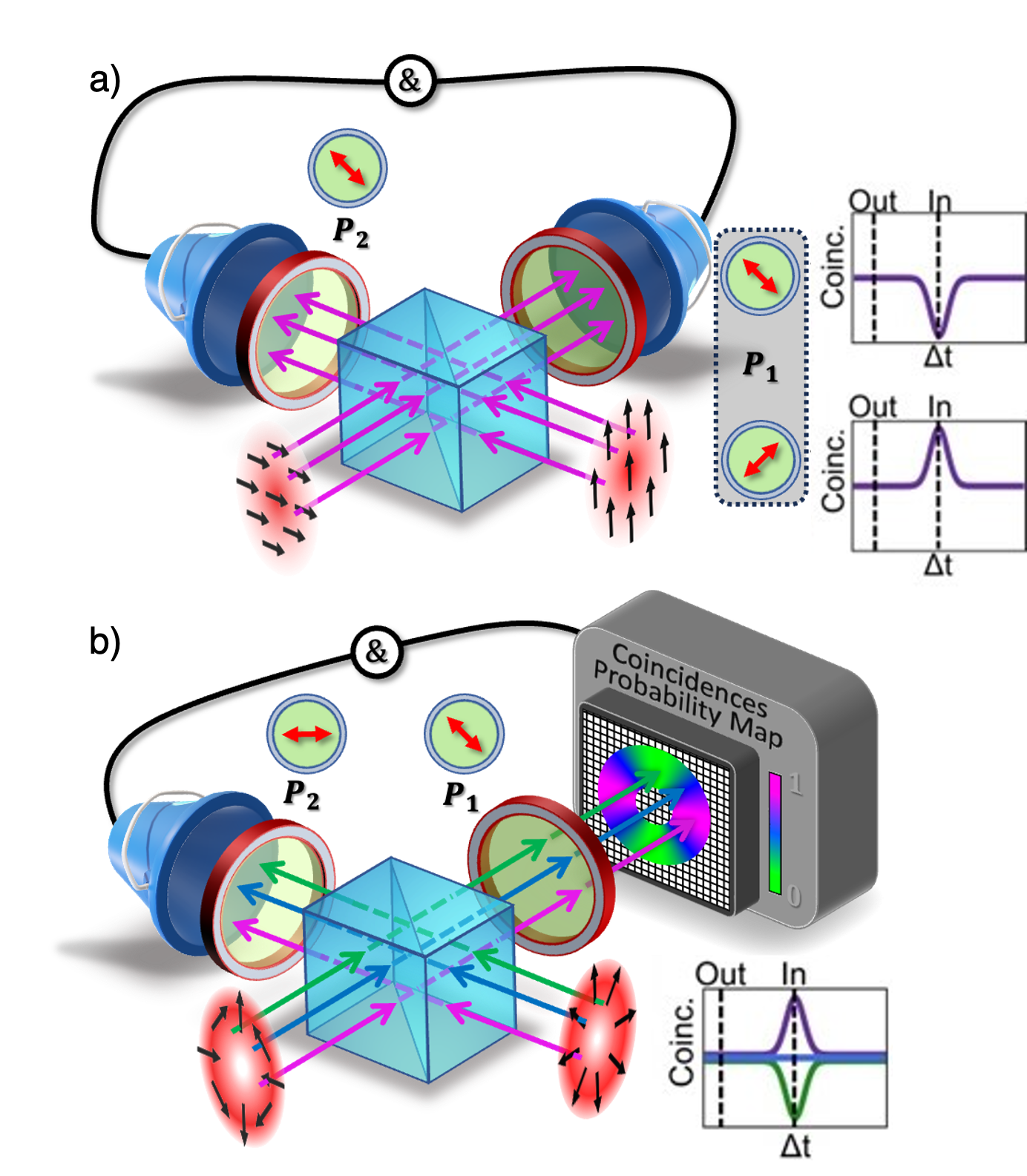}
\caption{\textbf{Spatially engineering quantum interference.} (a) In a quantum eraser, two photons in different polarisation states are quantum interfering when their \textit{which-path} information is erased through two polarisers. Depending on the polariser orientation, it is possible to tune photon coalescence from bunching to antibunching. 
This is reflected in a HOM curve (when using bucket detection of coincidences) that shows a dip or peak when the photons are temporally indistinguishable (in), relative to the coincidence rate measured when temporally distinguishable (out). (b) The polarisation distribution of input photons can be spatially tailored to engineer local variations in the quantum interference when overlapped on a 50:50 BS.  The interference space-variant behaviour is fully illustrated for three representative specific locations, depicted using distinct colours (blue, green, pink). When tuning the temporal distinguishability, the polarisation mixing in the quantum eraser results in a corresponding space-variant distribution of HOM, exhibiting peaks, dips or flat curves. This takes place despite one photon being detected with a bucket detector.}
\label{concept}
\end{figure}

\section{Results}

\noindent \textbf{Concept.} The concept behind structuring quantum interference for tailoring photon coalescence is illustrated in Figure \ref{concept}. Although typical HOM scenarios imply identical photons, quantum interference can be obtained also when the two photons entering the BS are distinguishable, thanks to the quantum erasing process \cite{Pan12}. When the \textit{which-path} information is encoded in polarisation, for instance, quantum interference can be enabled by placing a polariser on each of the output paths, thus performing two projective measurements that erase the \textit{which-path} information. Interestingly, depending on the initial polarisation state and each polariser's orientation, it is possible to fully tune photon coalescence from bunching to antibunching [Fig. \ref{concept}(a)] \cite{kwiat1992observation}. 
We could therefore exploit this feature in order to generate a structured quantum state with a tailored coalescence in the transverse plane if we design a spatially dependent quantum eraser. To this end, we consider a scenario where the polarisation of each, otherwise indistinguishable photon, is given a non-uniform distribution in the transverse plane before impinging on separate ports of a 50:50 BS. A polariser is moreover placed in each of the output paths of the BS. By tailoring the polarisation profile of each photon and selecting the polariser's orientation, we can have full control over the spatial distribution of the photonic coalescence [Fig. \ref{concept}(b)]. 
In other words, by structuring quantum interference, photon bunching is tailored along the optical mode transverse profile so that in each position we can have a zero-or-two photon NOON state (HOM dips), a single photon (HOM peak), or some combination of the two. 
\\

\begin{figure*}[t!]
\centering
\includegraphics[width=0.9\linewidth,keepaspectratio=true]{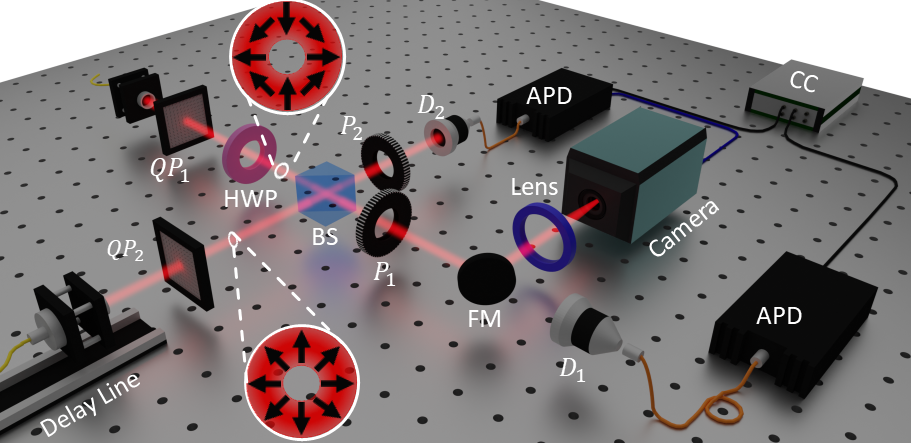}
\caption{\textbf{Experimental setup.}  After the spatial filtering with single mode fibers, two photons ($\lambda = 810$ nm), one in each path, are sent into two different input ports of a 50:50 beamsplitter (BS) as in a typical HOM setting. However, before reaching the BS, each of the photons is prepared in the desired vector vortex (VV) mode (with spatial profiles shown as insets) by sending horizontally polarised light through \textit{q}-plates (QP) with $q=0.5$. Placement of an additional half-waveplate (HWP) after QP$_1$ transformed one of the photons to the $\pi$ mode. The temporal delay between the two photons was controlled by a translation stage mounted in the arm containing QP$_2$. To realise bucket detection, the output photons were coupled (D$_1$,D$_2$) into multimode fibres and sent to two avalanche photodiodes (APD) connected to a coincidence counter (CC). For spatially resolved detection, one output was redirected via a flip mirror (FM) and focused with a lens onto an intensified charge coupled device (Camera), and gated upon detection of the other photon (D$_2$). \textit{Which-way} projections were performed by rotating linear polarisers $P_1$ and $P_2$.
}
\label{setup.png}
\end{figure*}

\noindent \textbf{Experimental Implementation.}
We experimentally demonstrate this concept by using the setup shown in Fig. \ref{setup.png}, with full details provided in Methods. Initially, SPDC biphotons ($\lambda = 810$ nm) are separated in path and spatially filtered with a single mode fiber. A set of polarisation correcting waveplates (not shown in Fig. \ref{setup.png}) facilitates uniform control of each photon's polarisation before being directed to the input ports of a 50:50 BS for interference, giving the two-photon state $\ket{\psi}_{in} = \left[\hat{a}_{A, m}^{\dag}(\tau) \hat{a}_{B, n}^{\dag}(\tau ')\right]\ket{0},\quad$ where $\hat{a}^{\dag}$ is the creation operator for the respective modes, ($n$,$m$), with the time-bins $\tau$ and $\tau'$, in the discrete paths $\{A, B\}$ of each input port. A motorised path delay is placed in one arm for tuning the temporal matching ($\tau' \rightarrow \tau$) of the photons. Initial matching of the input polarisation replicates the standard HOM experiment when tuning the delay. Further insertion of geometric phase elements, known as \textit{q}-plates (QP) \cite{qplate}, before the BS allows the controlled fashioning of spatially varying polarisation in each arm and the tuning of each photon's incident polarisation state, dependent on the topological charge (\textit{q}) of the element. 
 This allows one to prepare a transverse mode ($n$) of the form, 
\begin{equation}
    \mathbf{E}_{n}(\mathbf{r}, t) = \mathbf{e}_{s}(\varphi) f(r, t - \tau)e^{-i \omega t},
    \label{Eq:mode}
\end{equation}
for each photon, where $f$ indicates the amplitude profile generally dependent on the radial ($r$) coordinate, $\{\mathbf{e}_{s}(\varphi)\}$ are the transverse unit vectors specifying the local mode polarisation as a function of the azimuthal angle $\varphi$, $\omega$ is the mean (carrier) frequency of the mode and time is $t$.

In particular, without loss of generality, we consider two modes that present the same polarisation in some regions of the transverse profile, as depicted in Fig. \ref{concept}(b). As such, we choose the \textit{radial} ($q=0.5$) and $\pi$ [$q=0.5$ followed by a half-waveplate (HWP)] modes depicted there, which are two first order orthogonal VV modes \cite{DAmb16}, to be the input states of the BS. These may respectively be described by the azimuthally dependent polarisations,
\begin{equation}
\begin{aligned}
    &
    \mathbf{e}_{rad}(\varphi) = \mathbf{e}_{H}\cos{\varphi} + \mathbf{e}_{V}\sin{\varphi}
\\
    &
    \mathbf{e}_{\pi}(\varphi) = \mathbf{e}_{H}\cos{\varphi} - \mathbf{e}_{V}\sin{\varphi},
\label{eq: unit-vector for vector vortex}
\end{aligned}
\end{equation}
where $e_H$ and $e_V$ are, respectively, the horizontal and vertical polarisation unit-vectors.

As in typical HOM experiments, the interference or coalescent behaviour of our modes are observed through the two-fold coincidences measured on the state 
\begin{equation} 
    \ket{\psi}_{out} = 
    \frac{1}{2} \left[ \hat{a}_{A, rad}^{\dag}(\tau) \hat{a}_{B, \pi}^{\dag}(\tau ') - \hat{a}_{A, \pi}^{\dag}(\tau ') \hat{a}_{B, rad}^{\dag}(\tau) \right]
    \ket{0},
\label{eq: two-photons state}
\end{equation}
corresponding to the  projection of the full BS-output state onto the subspace in which the
two photons are separated in ports A and B, and may be calculated by the fourth-order correlation function \cite{Intro_QED}, as detailed in the Supplementary Information (SI). With this, we find the spatially varying coincidence probabilities for arbitrary polarisation projections \{$\alpha$,$\beta$\} on the photons detected in each port (respectively corresponding to coordinates $\varphi_1$ and $\varphi_2$) when temporally tuned ($\tau = \tau'$) 
\begin{equation}
\centering
\begin{split}
    C_{(In)}^{\alpha, \beta}(\varphi_1; \varphi_2) = 
    \\ &
    \frac{1}{4} 
    \left\vert
    \left[ \mathbf{u}_{\alpha} \cdot \mathbf{e}_{rad}(\varphi_{1}) \right]
    \left[ \mathbf{u}_{\beta} \cdot \mathbf{e}_{\pi}(\varphi_{2}) \right] -
    \right.
    \\ &
    \left.
    \left[ \mathbf{u}_{\alpha} \cdot \mathbf{e}_{\pi}(\varphi_{1}) \right]
    \left[ \mathbf{u}_{\beta} \cdot \mathbf{e}_{rad}(\varphi_{2}) \right]
    \right\vert^2
    \quad ,
\end{split}
\label{eq: coincidence prob - in - projected}
\end{equation}
and temporally distinguishable ($\tau \neq \tau'$) 
\begin{equation}
\centering
\begin{split}
    C_{(Out)}^{\alpha, \beta}&(\varphi_{1}; \varphi_{2}) = 
    \\&
    \frac{1}{4}
    \{
    \left\vert
    \left[ \mathbf{u}_{\alpha} \cdot \mathbf{e}_{rad}(\varphi_{1}) \right]
    \left[ \mathbf{u}_{\beta} \cdot \mathbf{e}_{\pi}(\varphi_{2}) \right]
    \right\vert^2
    +
    \\ &
    \left\vert
    \left[ \mathbf{u}_{\alpha} \cdot \mathbf{e}_{\pi}(\varphi_{1}) \right]
    \left[ \mathbf{u}_{\beta} \cdot \mathbf{e}_{rad}(\varphi_{2}) \right]
    \right\vert^2
    \}
    \quad .
\end{split}
\label{eq: coincidence prob - out - projected}
\end{equation}
Here $\mathbf{u}_{\alpha}$ and $\mathbf{u}_{\beta}$ are the unit-vectors directed along the detection polariser axes.
For simplicity, we ignore here the radial distribution as the polarisation variation depends only on $\varphi$ (see SI for a more general treatment). Experimentally, we realise these polarisation state projections by respectively placing polarisers $P_1$ and $P_2$ in each output port of the BS.

As depicted in Figure \ref{setup.png}, we adopted two strategies to observe the generated structure by means of these coincidences, both in and out of the temporal tuning. In one case, the photons from each output port were collected by coupling into multimode fibres connected to bucket detectors placed in each arm (flip mirror down), rendering no resolution of the spatial distribution. This corresponds to measuring the correlations given in Eqs. (4) and (5) after integrating over both azimuthal angles $\varphi_1$ and $\varphi_2$. In the second case (flip mirror up), we performed spatially-resolved measurements of one photon by replacing one bucket detector with a camera, which was then conditioned on the spatially-unresolved detection of another photon by the bucket detector in the other arm. This corresponds to measuring the correlations given in Eqs. (4) and (5) after integrating over one angle only, e.g. $\varphi_2$.

\begin{figure*}[ht]
\centering
\includegraphics[width=0.9\linewidth,keepaspectratio=true]{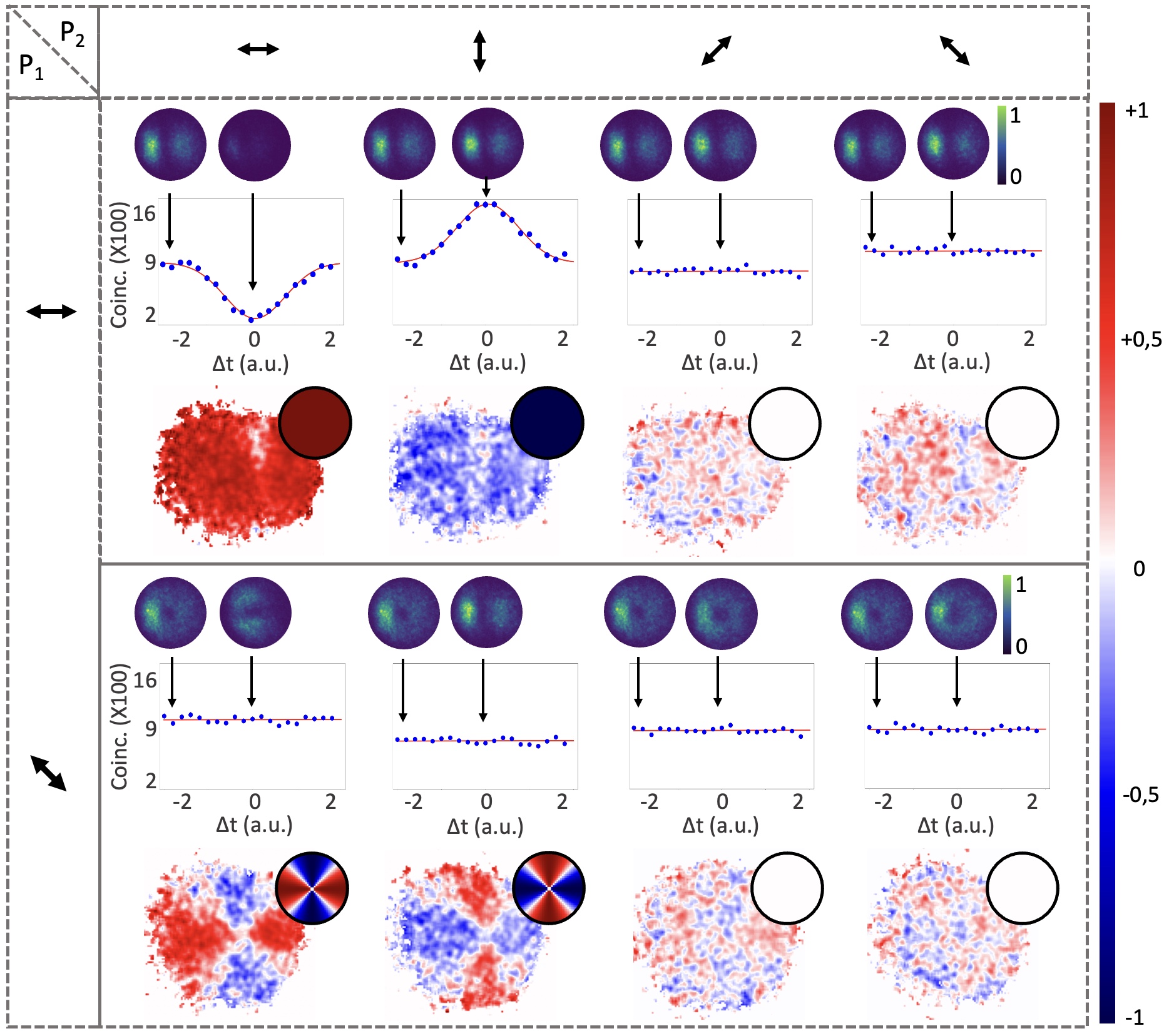}
\caption{\textbf{ Spatially tailored photon distributions from quantum interference.} 
Table of experimental outcomes for different choices of the polarisation projections P1 (black arrows, row-wise) and P2 (black arrows, column-wise). Hong-Ou-Mandel coincidence counts measured without spatial resolution, varying the temporal delay, are shown as blue points in the graphs in the upper section of each panel (solid curves are corresponding best fits). Error bars from Poissonian statistics are too small to observe. The related spatially-resolved coincidence images (spanning $\sim$ 60 X 60 pixels) in and out of the temporal distinguishability, are given as insets to the HOM curves with arrows denoting the measured distribution for the temporally tuned ($\Delta t = 0$, \enquote*{In}) case to the right and temporally distinguishable ($\Delta t <<0$, \enquote*{Out}) case to the left. The related spatially resolved visibilities are given as the transverse distributions on the bottom, using the false-color scale shown on the right-side bar. Analytically calculated visibility distributions are given as top-right circular insets. 
} 
\label{fig: Experimental Results}
\end{figure*}

It is now possible to identify polarisation projection measurements that erase the \textit{which-path} information and enable quantum interference between these structured modes, even for bucket detection. 
Intuitively, one may consider the VV spatial distributions depicted in Fig. \ref{setup.png} which shows they share the same polarisation state along the horizontal (H) and vertical (V) axes, while being completely orthogonal along the diagonal (D) and anti-diagonal (A) directions. 
It thus follows that a projection along H and V should destroy the \textit{which-path} information for the latter pairing and recover HOM interference. 
To observe the related coalescence for such polarisations ($\alpha$ and $\beta$) we define the visibility,  $\mathcal{V}_{\alpha,\beta}=(C^{\alpha,\beta}_{out}-C^{\alpha,\beta}_{in})/ C^{\alpha,\beta}_{out} \in [-1,1]$ between the coincidences detected out ($C_{out}$) and in ($C_{in}$) the temporal indistinguishably criteria [as depicted in Fig. \ref{concept} (b)]. 
This allows us to easily recognise bunching as $\mathcal{V}$ becomes positive from anti-bunching where $\mathcal{V}$ becomes negative with extreme cases of perfect bunching (antibunching) being detected when $\mathcal{V} = 1$  $(-1)$.  

The graphs given in each panel of Figure \ref{fig: Experimental Results} show the experimental outcomes of the bucket-only detection as the delay between the two input photons was varied by moving the delay line.
Here, projections $(P_1,P_2)=$ (H,H) elicit a dip at $\Delta t = 0$, coinciding with a erasure of the which-way information for the two photons and thereby restoring quantum interference.
For $(P_1,P_2)=$ (H,V), we find a peak instead of a dip, corresponding to photon antibunching. \cite{kwiat1992observation,Walb02}
Alternatively, cases $(P_1,P_2)=$ (H,A),(H,D),(A,D) and (A,A) result in a flat curve (no HOM dip nor peak).

On the other hand, for spatially-resolvable measurements (using the camera), we expect the following spatially structured visibilities, corresponding to a set of polarisation projections of $P_1 \in \{H,A\}$ in the camera arm and $P_2 \in \{H, V, D, A\}$ in the bucket detector arm as depicted by the arrows in Fig. \ref{fig: Experimental Results}: 
\begin{equation}
\begin{aligned}
&
\mathcal{V}_{HH} = 1
\\
&
\mathcal{V}_{HV} = -1
\\
&
\mathcal{V}_{AH} = \cos{(2\varphi_1)}
\\
&
\mathcal{V}_{AV} = - \cos{(2\varphi_1)} \\
&
\mathcal{V}_{HA} = \mathcal{V}_{HD} = \mathcal{V}_{AA} = \mathcal{V}_{AD} = 0,
\end{aligned}
\label{Eq:StructuredVis}
\end{equation}

obtained by integrating (\ref{eq: coincidence prob - in - projected}) and (\ref{eq: coincidence prob - out - projected}) over $\varphi_{2}$ due to bucket detection in arm 2.
In this case, images conditioned on coincidence with the bucket detector were captured \enquote{out} (upper left inset of the curves) and \enquote{in} (upper right inset of the curve) the temporally matched condition. Some asymmetry towards the left in the distributions may be noted as the result of experimental misalignment in the delay line to the camera (see Methods) as well as the $q$-plate with respect to the photon distribution. This, however, does not significantly detract from the expected distributions. Each measured pair of coincidence distributions then allowed us to reconstruct the visibility map (lower row) for the projection settings. 

The resulting outcomes relate well to what was observed in the case of no spatial resolution (bucket detection) as the coincidences measured by bucket detection are proportional to the integrated pixel distribution over the mode profile. We additionally compare these to the calculated profiles from Eq. \ref{Eq:StructuredVis} that are shown as upper-right insets of the visibility maps.
Accordingly, $(P_1,P_2)=$ (H,H) shows uniform coalescence across the spatial distribution, while (H,V) exhibits a uniform anti-coalescence, within the expectations of the calculated distributions. 
The cases of HA, HD, AA and AD similarly yield a uniform structure in line with expectations of a zero visibility across the measured profile.

Crucially, projections AH and AV,  reveal spatially varying features with two complementary lobes, showing coalescence and anti-coalescence across the structure as predicted by Eqs. \ref{Eq:StructuredVis}.  
These structures are not detectable without spatial resolution, as shown by the flat curves in the corresponding graphs.
As a result, we herald a mode of structured bosonic coalescence with some parts containing either zero or two photons as in the case of NOON states. At the same time, other regions of the mode, corresponding to anti-coalescence,  will always be populated by one photon. 

\noindent \textbf{Discussion.} 
With these results, we find interesting features, most notable are the asymmetries in projected states made in either arm. For instance, a diagonal projection only reveals structurally varying coalescence when placed before the camera and one only observes total non-zero visibilities for H and V polarisations projections. To understand this, one must consider, not just the polarisation, but the paired spatial structures associated with them as well as the global distinguishability. Here, projection on H and V for both VV modes results in identical spatial lobes that overlap for both distributions, thus rendering the photons indistinguishable, both globally and locally. For A and D, however, the lobes, while partially overlapping, are orthogonal structures and thus globally distinguishable. It follows that if one were to integrate over the first case, the structure is indistinguishable and thus facilitates in erasing the which-way information, while in the second case the distinguishability is held. As the photons must be indistinguishable in both arms to erase the \textit{which-path} information and observe interference, the perceived distinguishability for both detectors determines what is observed. Bucket detection with A and D projections thus remains distinguishable, leading to no visibility, even with an indistinguishable projection on the camera. Conversely, bucket detection with H and V allows the locally indistinguishable structures for A and D to be seen, despite the global distinguishability.
Additionally, a distinct reversal of the coalescent and anticoalescent distributions can be seen for complementary polarisation projections when structural coalescence or anticoalescence is present (i.e. $P_2 =$ H vs V for $P_1 =$ H and A).  One can thus locally switch between bosonic and fermionic behaviour, or equivalently, between two/zero (NOON) and single photon occupation.   
It follows that tailoring the quantum state can be conveniently achieved by changing the polarisation distribution of the input beams as well as the polarisation projections.

In conclusion, we have generated structured quantum states of light that have no classical counterpart, with a spatially tailored bosonic coalescence. 
We achieved this by exploiting a simple and robust scheme based on spatially tailored distinguishability between two photons in a quantum eraser setting. 
This additionally invites insight into the structured state where, in particular, the use of orthogonal modes for each interfered photon showed a global versus local HOM effect \cite{Nits20} such that detection without spatial resolution yielded no sign of interference, but spatially resolved measurements reveals its presence. 
Although these results were demonstrated for VV modes, our technique  may be generalised to intelligently engineer arbitrary desired interference structures for varying purposes.
Additionally, we point out that 
programmable approaches, for tailoring the input polarisation distributions, \cite{liu2022flexible} would allow dynamic on-demand variations of these states. As distinguishabilty is the property one needs to tailor, one may extend our result to other degrees of freedom, such as temporally structured light \cite{Spec09}. 
By nature of the ability to spatially tailor the coalescent behaviour of photons, we believe these new states and engineering thereof could be beneficial for fields such as quantum microscopy, metrology and communications, including high-dimensional protocols.  \\

\section{Methods}
\noindent \textbf{Experimental details.} The two ($\lambda = 810$ nm) signal-idler photons used in the experiment (Fig. \ref{setup.png}) for interference on the BS were generated in the state $\ket{\psi}=\ket{H}\ket{V}$ via type-II degenerate collinear SPDC from a 30 mm long ppKTP crystal pumped with a $\lambda = 405$ nm continuous wave laser and temperature phase-matched at $\sim$ 38 $^\circ$C.
They were then path separated with a PBS and each spatially filtered by coupling into single mode fibers (SMF) which directed the photons through a half-waveplate and quater-waveplate (for preparation into the horizontal polarisation state) as well as electronically tuned \textit{q}-plates with the topological charge $q=0.5$ for polarisation-dependent spatial structuring,  before impinging onto a 50:50 BS for interference. One SMF was placed on a motorised translation stage to introduce a path-delay for temporally tuning the distinguishability. The output of each BS port was then directed towards detectors. We switched between space-integrated and space-resolved measurements by using a flip mirror. The first case (flip mirror down) coupled the structured photons in each arm into multimode fibres for detection by 
single-photon counting modules (APD) connected to a 
coincidence counter with an integration window of 2 ns. The second case saw the photons in one arm diverted by the flip mirror to a 30 m delay line for compensation of the electronic delay associated with the triggering of the camera, before being focused by a $f = 150$ mm lens onto an intensified CCD. 
This camera was then gated by the trigger provided from the APD, positioned in the other arm, so as to spatially capture the photons in coincidence. Polarisation projections on the structured states were made by inserting linear polarisers into each output arm before the relevant detectors.\\

\noindent \textbf{Vector vortex modes.} In the experiment, VV modes were generated by exploiting spin-to-orbital angular momentum conversion in a birefringent liquid crystal slab with uniform retardation and an azimuthally varying optical axis, known as a \textit{q}-plate (QP). This device is chatacterised by a topological charge, $q$, related to the number of rotations of the local optical axis and, accordingly, the polarisation-dependent geometric phase imparted to incident light. In Jones matrix formalism, the QP operation in the linear basis, where $\ket{H} = [1;0]^T$ and $\ket{V} = [0;1]^T$ are horizontal and vertical polarisation states, is described as,
\begin{equation}
    QP = \begin{bmatrix}
        \cos{}(2q\varphi)& \sin{}(2q\varphi)\\
        \sin{}(2q\varphi)& -\cos{(2q\varphi)}
    \end{bmatrix}.
\end{equation}
From this, it is straightforward to see the spatially-varying polarisation distribution induced and dependence on polarisation. Taking a horizontally polarised state as in the experiment, one obtains the following output:
\begin{equation}
\begin{aligned}
    QP\ket{H} &= \begin{bmatrix}
        \cos{}(2q\varphi)& \sin{}(2q\varphi)\\
        \sin{}(2q\varphi)& -\cos{}(2q\varphi)
    \end{bmatrix}
    \begin{bmatrix}
        1 \\
        0
    \end{bmatrix}
    =\begin{bmatrix}
        \cos{}(2q\varphi)\\
        \sin{}(2q\varphi)
    \end{bmatrix}
\end{aligned}
\end{equation}
and with $q=0.5$ as in the experiment, the state described in Eq. \ref{eq: unit-vector for vector vortex} is obtained. By adding a HWP = $[1, 0; 0, -1]^T$ where the optical axis aligned horizontally with the QP, the following transformation is achieved:
\begin{equation}
\begin{aligned}
    [HWP][QP] &=\begin{bmatrix}
        1 & 0\\
        0 & -1
    \end{bmatrix}
    \begin{bmatrix}
        \cos{}(2q\varphi)& \sin{}(2q\varphi)\\
        \sin{}(2q\varphi)& -\cos{}(2q\varphi)
    \end{bmatrix}\\
    &= \begin{bmatrix}
        \cos{}(2q\varphi)& \sin{}(2q\varphi)\\
        -\sin{}(2q\varphi)& \cos{}(2q\varphi)
    \end{bmatrix}
\end{aligned}
\end{equation}

that, acting on the state $\ket{H}$, produces the desired state $\ket{\pi} = \cos{(\varphi)}\ket{H} -\sin{(\varphi)}\ket{V}$ for $q = 0.5$. 

\section*{Acknowledgements}
We acknowledge support from the Italian Ministry of Research (MUR) through the PRIN 2017 project “Interacting photons in polariton circuits” (INPhoPOL) and the PNRR project PE0000023-NQSTI. We also acknowledge support from NATO through SPS Project HADES - MYP G5839 and from the Italian Space Agency (ASI) through the "High dimensional quantum information" (HDQI) project.
\bibliographystyle{naturemag}
\bibliography{biblio.bib}
\end{document}


\title{Supplementary Information: Engineering quantum states from a spatially structured quantum eraser}

\author{Carlo Schiano}
\affiliation{Dipartimento di Fisica, Universit\`a di Napoli Federico II, Complesso Universitario di Monte S. Angelo, Via Cintia, 80126 Napoli, Italy}
\author{Bereneice Sephton}
\affiliation{Dipartimento di Fisica, Universit\`a di Napoli Federico II, Complesso Universitario di Monte S. Angelo, Via Cintia, 80126 Napoli, Italy}
\author{Roberto Aiello}
\affiliation{Dipartimento di Fisica, Universit\`a di Napoli Federico II, Complesso Universitario di Monte S. Angelo, Via Cintia, 80126 Napoli, Italy}
\author{Francesco Graffitti}
\affiliation{Institute of Photonics and Quantum Sciences, School of Engineering and Physical Sciences, Heriot-Watt University, Edinburgh EH14 4AS, UK}
\author{Nijil Lal}
\affiliation{Dipartimento di Fisica, Universit\`a di Napoli Federico II, Complesso Universitario di Monte S. Angelo, Via Cintia, 80126 Napoli, Italy}
\author{Andrea Chiuri}
\affiliation{Enea - Centro Ricerche Frascati, via E. Fermi 45, 00044 Frascati, Italy}
\author{Simone Santoro}
\affiliation{Enea - Centro Ricerche Frascati, via E. Fermi 45, 00044 Frascati, Italy}
\author{Luigi Santamaria Amato}
\affiliation{Italian Space Agency (ASI), Centro di Geodesia Spaziale ‘Giuseppe Colombo’, Località Terlecchia, 75100 Matera, Italy}
\author{Lorenzo Marrucci}
\affiliation{Dipartimento di Fisica, Universit\`a di Napoli Federico II, Complesso Universitario di Monte S. Angelo, Via Cintia, 80126 Napoli, Italy}
\author{Corrado de Lisio}
\affiliation{Dipartimento di Fisica, Universit\`a di Napoli Federico II, Complesso Universitario di Monte S. Angelo, Via Cintia, 80126 Napoli, Italy}
\author{Vincenzo D'Ambrosio}
\email[]{vincenzo.dambrosio@unina.it}
\affiliation{Dipartimento di Fisica, Universit\`a di Napoli Federico II, Complesso Universitario di Monte S. Angelo, Via Cintia, 80126 Napoli, Italy}

\maketitle
\section{Structural correlation formulation}
We send two structured photon beams into a beamsplitter (BS) and measure two-fold coincidence between a bucket detector and a camera, after polarisation projection.
We record different spatial structures on the camera, depending on the polarisation projection performed and on the choice of the input structured beams.
Here we describe in detail how it works with theoretical calculations.

According to the quantum theory of electrodynamics, the transverse electric field operator can be written as \cite{Photons_Atoms}:
\begin{equation}
    \hat{\mathbf{E}}_{\perp}(\mathbf{r}, t) = \hat{\mathbf{E}}_{\perp}^{+}(\mathbf{r}, t) + \hat{\mathbf{E}}_{\perp}^{-}(\mathbf{r}, t)
\end{equation}
where 
$\hat{\mathbf{E}}_{\perp}^{+}(\mathbf{r}, t) = \sum_{n} \mathbf{E}_{n}(\mathbf{r}, t) \hat{a}_{n}$ 
and 
$\hat{\mathbf{E}}_{\perp}^{-}(\mathbf{r}, t) = \sum_{n} \mathbf{E}_{n}^{*}(\mathbf{r}, t) \hat{a}_{n}^{\dag}$ 
are the positive and negative frequencies field components, respectively.
The index $n$ spans a complete set of normal modes $\{\mathbf{E}_{n}(\mathbf{r}, t)\}$, and $\hat{a}_{n}$ and $\hat{a}_{n}^{\dag}$ are the photon annihilation and creation operators, respectively, with the usual commutation rules $\left[ \hat{a}_{n}, \hat{a}_{n'}^{\dag}\right] = \delta_{n, n'}$.
We are here assuming for convenience that these modes fill a finite space volume $\Omega$ and satisfy suitable boundary conditions at the volume border so as to discretize the volume set.

For any given quantum state $\ket{\psi}$ (we assume pure quantum states here, but the treatment can be easily generalized to the general case of quantum mixtures), the number of double detections per unit of time between polarised photons hitting two separate broad-band pointlike detector is proportional to the fourth-order correlation function \cite{Intro_QED}:
\begin{equation}
    w_{II}^{\alpha, \beta}(\mathbf{r}_{1}, t_{1}; \mathbf{r}_{2}, t_{2}) = 
    \bra{\psi}
    \left[\mathbf{u}_{\alpha}\cdot\hat{\mathbf{E}}_{\perp}^{-}(\mathbf{r}_{1}, t_{1})\right]  
    \left[\mathbf{u}_{\beta}\cdot\hat{\mathbf{E}}_{\perp}^{-}(\mathbf{r}_{2}, t_{2})\right] 
    \left[\mathbf{u}_{\beta}\cdot\hat{\mathbf{E}}_{\perp}^{+}(\mathbf{r}_{2}, t_{2})\right]  
    \left[\mathbf{u}_{\alpha}\cdot\hat{\mathbf{E}}_{\perp}^{+}(\mathbf{r}_{1}, t_{1})\right]
    \ket{\psi}
\label{eq: second order correlations - polarizer}
\end{equation}
where $\mathbf{u}_{\alpha}$ and $\mathbf{u}_{\beta}$ are the unit-vectors corresponding to the polarizations being detected at $r_1$ and $r_2$, respectively.

\begin{figure}[ht]
\centering
\begin{subfigure}[b]{0.4\textwidth}
  \centering
  \includegraphics[width=\textwidth]{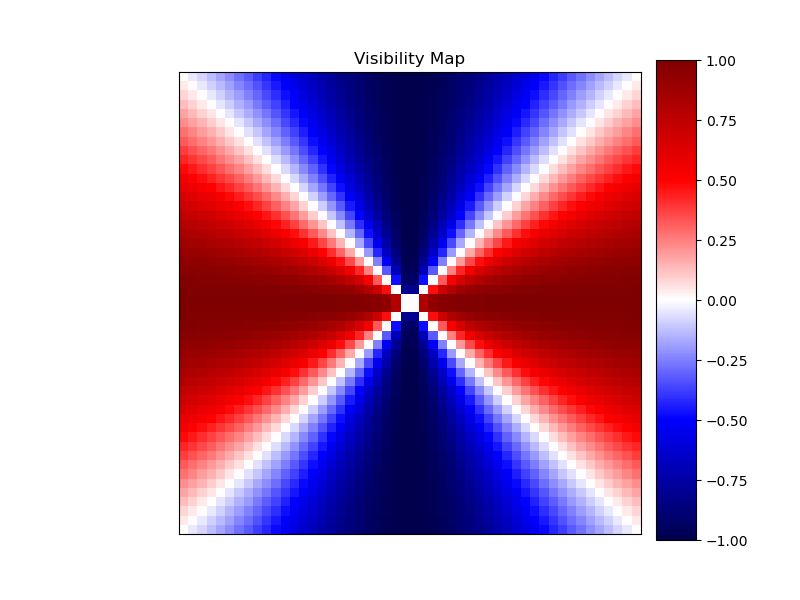}
  \caption{A-H}
\end{subfigure}
\hspace{2pt}
\begin{subfigure}[b]{0.4\textwidth}
  \centering
  \includegraphics[width=\textwidth]{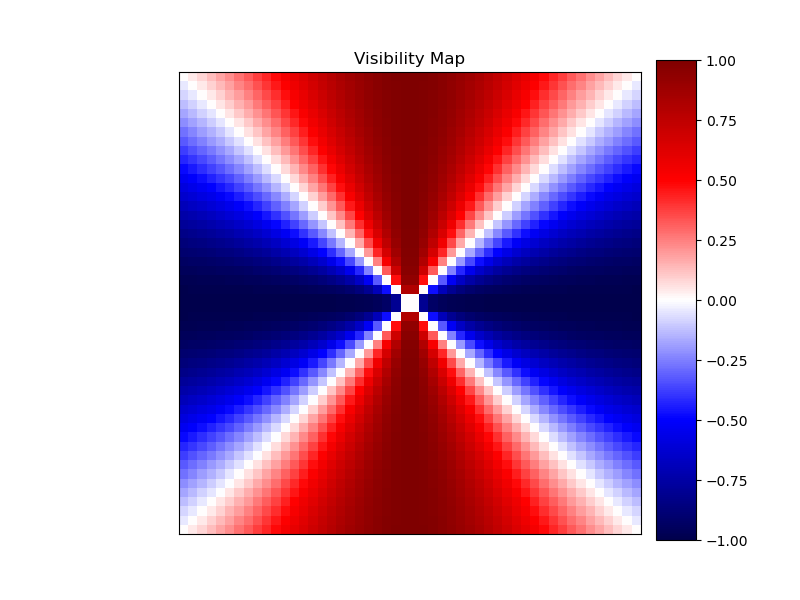}
  \caption{A-V}
\end{subfigure}
\caption{Visibility distribution for A-H and A-V polarisers configurations.}
\label{fig: Visibility Maps}
\end{figure}

For our purposes, it is convenient to choose a set of modes that are eigenstate of vector beams, described by the OAM eigenvalue $m$ ($L_{z} = m \hbar$) and by  a non-negative integer $p$ specifying different orthogonal radial modes for each $\abs{m}$. 
In our case, we consider quasi-monochromatic pulsed modes with mean (carrier) frequency $\omega$, belonging to discrete paths $P\in\{A,B\}$ and having a specified time bin $\tau$ (so that delay can be used to distinguish the photons).
In the paraxial approximation, assuming that $z$ is the main propagation axis, these modes can be written as follows:
\begin{equation}
    \mathbf{E}_{n}(\mathbf{r}, t) = \mathbf{e}_{s} f_{\abs{m}, p}(r_{P}, z_{P}, t - \tau)e^{im\varphi_{P}}e^{-i \omega t}
\label{eq: final field mode}
\end{equation}
where $(r_{P}$, $z_{P}$, $\varphi_{P})$ are cylindrical coordinates for the given path $P$, $f_{\abs{m}, p}$ specifies the radial profile of the mode amplitude as a function of $z$, and $\{\mathbf{e}_{s}\}$ are
two unit-vectors orthogonal to $z$ with polarisation index $s$. 
Thus, the mode index can be written as $n = (P, m, p, \omega, \tau, s)$.
We assume that different values of the various indices correspond to orthogonal modes.
Rather than describing the vector beams as superpositions of modes like (\ref{eq: final field mode}), it is also possible (and in our case simpler) to introduce in eq. (\ref{eq: final field mode}) an azimuthally-position-dependent polarisation, $\mathbf{e}_{s}(\varphi)$ (in this case the $m$ integer is not the OAM eigenvalue anymore). 
In our experiment, the vector beams impinging on the faces of the BS are the radial and $\pi$ beam (with total angular momentum 2 and OAM 1), that are respectively described by the following azimuthally-position-dependent polarisations
\begin{equation}
\begin{aligned}
    &
    \mathbf{e}_{rad}(\varphi) = \mathbf{e}_{H}\cos{\varphi} + \mathbf{e}_{V}\sin{\varphi}
    \\
    &
    \mathbf{e}_{\pi}(\varphi) = \mathbf{e}_{H}\cos{\varphi} - \mathbf{e}_{V}\sin{\varphi}
\end{aligned}
\end{equation}
where $e_{i}$, $i \in \{H,V\}$ are the horizontal ($H$) and vertical ($V$) polarisation unit-vector.
Let us now consider certain more specific states, obtained by having two photons in well-defined vector beams (vv1 and vv2) impinging on the two input faces of a beamsplitter (BS). 
We assume that all properties of the photon beams, except their time bins, are identical (we omit them for brevity), so that detected photons become indistinguishable only if $\tau = \tau '$. 
By making these assumptions, the quantum state $\ket{\psi}$ in eq. (\ref{eq: second order correlations - polarizer}) can be obtained by acting with a 50:50 beamsplitter on the two-photon state $\ket{\psi}_{in} = a^{\dag}_{A,vv1}(\tau)a^{\dag}_{B,vv2}(\tau ')\ket{0}$ and thus it is given by
\begin{equation}
    \ket{\Psi} = \frac{1}{2} 
    \left( \hat{a}_{A, vv1}^{\dag}(\tau) + i \hat{a}_{B, vv1}^{\dag}(\tau) \right) 
    \left( i \hat{a}_{A, vv2}^{\dag}(\tau ') + i \hat{a}_{B, vv2}^{\dag}(\tau ') \right) \ket{0}
\label{eq: two-photons state}
\end{equation}
where $a^{\dag}$ is the creation operator for the $vv1$ and $vv2$ modes, with the time-bins, $\tau$ and $\tau '$, in the discrete paths $\{A,B\}$ of each BS port.
Because we are only interested in the coincidences obtained when placing one detector in path A and the
other in path B, we may now project (or “postselect”) this state within the subspace in which the
two photons are separated in ports A and B (the other cases do not give coincidences). 
We thus obtain the following expression for the unnormalized (projected) quantum state
\begin{equation} 
    \ket{\Psi} =
    \frac{1}{2} \left[ \hat{a}_{A, vv1}^{\dag}(\tau) \hat{a}_{B, vv2 '}^{\dag}(\tau) - \hat{a}_{A, vv2}^{\dag}(\tau ') \hat{a}_{B, vv1}^{\dag}(\tau) \right]
\label{eq: two-photons state - final}
\end{equation}
that corresponds to the two-photon input state of eq. (3) in the main text (if we put $vv1=rad$ and $vv2=\pi$). 
Inserting (\ref{eq: two-photons state - final}) into (\ref{eq: second order correlations - polarizer}) and exploiting the creation-annihilation operator commutation rules, we immediately obtain the expected fourth-order correlation function
\begin{equation}
\begin{split}
    w_{II}^{\alpha, \beta}(\mathbf{r}_{1}, t_{1}; \mathbf{r}_{2}, t_{2}) = 
    \frac{1}{4}
    \left|
    \left[ \mathbf{u}_{\alpha} \cdot \mathbf{E}_{A, vv1, \tau}(\mathbf{r}_{1}, t_{1}) \right] 
    \left[ \mathbf{u}_{\beta} \cdot \mathbf{E}_{B, vv2, \tau '}(\mathbf{r}_{2}, t_{2}) \right] - 
    \right. \\ \left.
    \left[ \mathbf{u}_{\alpha} \cdot \mathbf{E}_{A, vv2, \tau '}(\mathbf{r}_{1}, t_{1}) \right] 
    \left[ \mathbf{u}_{\beta} \cdot \mathbf{E}_{B, vv1, \tau}(\mathbf{r}_{2}, t_{2}) \right]
    \right|^{2}    
\end{split}
\label{eq: double detections rate - two photons state}
\end{equation}
If we now insert the modes expression (\ref{eq: final field mode}) into (\ref{eq: double detections rate - two photons state}), we obtain
\begin{equation} 
\begin{split}
    w_{II}^{\alpha, \beta}(\mathbf{r}_{1}, t_{1}; \mathbf{r}_{2}, t_{2}) = 
    \frac{1}{4} 
    \left|
    \left[ \mathbf{u}_{\alpha} \cdot \mathbf{e}_{vv1}(\varphi_{1})f(r_{1}, z_{1}, t_{1} - \tau) \right] 
    \left[ \mathbf{u}_{\beta} \cdot \mathbf{e}_{vv2}(\varphi_{2})f(r_{2}, z_{2}, t_{2} - \tau ') \right] - \right. \\
    \left.
    \left[ \mathbf{u}_{\alpha} \cdot \mathbf{e}_{vv2}(\varphi_{1})f(r_{1}, z_{1}, t_{1} - \tau ') \right]
    \left[ \mathbf{u}_{\beta} \cdot \mathbf{e}_{vv1}(\varphi_{1})f(r_{1}, z_{1}, t_{1} - \tau) \right]
    \right|^{2} \quad .
\end{split}
\label{eq: double detections rate - two photons state - OAM modes}
\end{equation}
In order to get the coincidence probability between two photons which polarisation are directed along $\mathbf{u}_{\alpha}$ and $\mathbf{u}_{\beta}$, this rate must be integrated both in $t_{1}$ and $t_{2}$ over the same detection time window
\begin{equation}
    C_{\alpha, \beta}(\mathbf{r}_{1}; \mathbf{r}_{2}) = \int w_{II}^{\alpha, \beta}(\mathbf{r}_{1}, t_{1}; \mathbf{r}_{2}, t_{2}) dt_{1} dt_{2} 
\end{equation} 
We must distinguish here the case of temporally tuned photons ($ \tau = \tau '$) from the case of temporally distinguishable photons ($\tau \neq \tau '$). 
In the former case, we are reproducing the "in the dip/peak" condition of a typical HOM experiment, and we obtain 
\begin{equation}
    C_{(In)}^{\alpha,\beta}(\mathbf{r}_{1}; \mathbf{r}_{2}) = 
    \frac{F(r_{1}, z_{1}) F(r_{2}, z_{2})}{4} 
    \abs{
    \left[ \mathbf{u}_{\alpha} \cdot \mathbf{e}_{vv1}(\varphi_{1}) \right]
    \left[ \mathbf{u}_{\beta} \cdot \mathbf{e}_{vv2}(\varphi_{2}) \right] -
    \left[ \mathbf{u}_{\alpha} \cdot \mathbf{e}_{vv2}(\varphi_{1}) \right]
    \left[ \mathbf{u}_{\beta} \cdot \mathbf{e}_{vv1}(\varphi_{2}) \right]
    }^2
\label{eq: C_in}
\end{equation}
while, in the latter case, assuming that both time bins fall in the integration window but do not overlap, we are reproducing the "out the dip/peak" condition of a typical HOM experiment, and we obtain
\begin{equation}
    C_{(Out)}^{\alpha,\beta}(\mathbf{r}_{1}; \mathbf{r}_{2}) = 
    \frac{F(r_{1}, z_{1}) F(r_{2}, z_{2})}{4}
    \left\{
    \abs{
    \left[ \mathbf{u}_{\alpha} \cdot \mathbf{e}_{vv1}(\varphi_{1}) \right]
    \left[ \mathbf{u}_{\beta} \cdot \mathbf{e}_{vv2}(\varphi_{2}) \right]
    }^2
    -
    \abs{
    \left[ \mathbf{u}_{\alpha} \cdot \mathbf{e}_{vv2}(\varphi_{1}) \right]
    \left[ \mathbf{u}_{\beta} \cdot \mathbf{e}_{vv1}(\varphi_{2}) \right]
    }^2
    \right\}
\label{eq: C_out}
\end{equation}
where we introduced the "fluence" function
\begin{equation}
    F(r, z) = \int\limits_{\substack{\text{integration} \\ \text{window}}} f^{2}(r, z, t - \tau)
\end{equation}
By setting $vv1=rad$ and $vv2=\pi$ and ignoring the radial distribution (as the polarisation variation depends only on $\varphi$), eqs. (\ref{eq: C_in}) and (\ref{eq: C_out}) reduce to eqs. (4) and (5) in main text, respectively.
In our experiment, the coincidence probability functions are measured by considering $vv1=rad$ and $vv2=\pi$ vector beams impinging the BS and by setting the polarisation $P_{1} \in \{H, A\}$ for the polariser located before the camera and $P_{2} \in \{H, V, A, D\}$ for the polariser located before the bucket detector, where $\hat{e}_{D} = (\hat{e}_{H} + \hat{e}_{V})/\sqrt{2}$ and $\hat{e}_{A} = (\hat{e}_{H} - \hat{e}_{V})/\sqrt{2}$ are the unit-vectors associated to D and A, respectively.
We are interested in theoretically evaluating the visibility map given by the formula
\begin{equation}
    \mathcal{V}_{\alpha,\beta}(\mathbf{r}_{1}; \mathbf{r}_{2}) = 
    \frac{
    C_{(Out)}^{\alpha,\beta}(\mathbf{r}_{1}; \mathbf{r}_{2}) - C_{(In)}^{\alpha,\beta}(\mathbf{r}_{1}; \mathbf{r}_{2})
    }
    {
    C_{(Out)}^{\alpha,\beta}(\mathbf{r}_{1}; \mathbf{r}_{2})  
    }
\end{equation}
The visibility map is calculated for the 8 different polarisation combinations listed above and the results are presented below.
We set $A_{1,2} = F(r_{1}, z_{1}) F(r_{2}, z_{2})$ in order to lighten up the notation in the following equations. 
In case of H-H and H-V polarisation settings, the visibility map does not depend on the azimuthal angles $\varphi_{1}$ and $\varphi_{2}$.
\begin{equation}
\left.
\begin{aligned}
&
C_{(Out)}^{H,H}(\mathbf{r}_{1}; \mathbf{r}_{2}) 
=
\frac{A_{1,2}}{2}
\cos^{2}{\varphi_{1}}\cos^{2}{\varphi_{2}}
\\
&
C_{(In)}^{H,H}(\mathbf{r}_{1}; \mathbf{r}_{2}) = 0
\end{aligned}
\right\rbrace \quad
\mathcal{V}_{HH}(\varphi_{1}; \varphi_{2}) = 1
\end{equation}
\begin{equation}
\left.
\begin{aligned}
&
C_{(Out)}^{H,V}(\mathbf{r}_{1}; \mathbf{r}_{2}) 
=
\frac{A_{1,2}}{2}
\cos^{2}{\varphi_{1}}\cos^{2}{\varphi_{2}}^{2}
\\
&
C_{(In)}^{H,V}(\mathbf{r}_{1}; \mathbf{r}_{2})
=
A_{1,2}
\cos^{2}{\varphi_{1}}\cos^{2}{\varphi_{2}}
\end{aligned}
\right\rbrace \quad
\mathcal{V}_{HV}(\varphi_{1}; \varphi_{2}) = -1
\end{equation}
In case of H-A, H-D, A-H, A-V polarisation settings, the visibility map depends only on the $\varphi_{2}$ angle for the H-A and H-D configurations, 
while it depends only on the $\varphi_{1}$ angle for the A-H and A-V configurations.
\begin{equation}
\left.
\begin{aligned}
&
C_{(Out)}^{H,A}(\mathbf{r}_{1}; \mathbf{r}_{2}) 
=
C_{(Out)}^{H,D}(\mathbf{r}_{1}; \mathbf{r}_{2}) 
=
\frac{A_{1,2}}{4}
\cos^{2}{\varphi_{1}}
\\
&
C_{(In)}^{H,A}(\mathbf{r}_{1}; \mathbf{r}_{2}) 
=
C_{(In)}^{H,D}(\mathbf{r}_{1}; \mathbf{r}_{2}) 
=
\frac{A_{1,2}}{2}
\cos^{2}{\varphi_{1}}\sin^{2}{\varphi_{2}}
\end{aligned}
\right\rbrace \quad
\mathcal{V}_{HA}(\varphi_{1}; \varphi_{2}) 
= 
\mathcal{V}_{HD}(\varphi_{1}; \varphi_{2}) 
= 
\cos{(2 \varphi_{2})}
\end{equation}
\begin{equation}
\left.
\begin{aligned}
&
C_{(Out)}^{A,H}(\mathbf{r}_{1}; \mathbf{r}_{2}) 
=
\frac{A_{1,2}}{4}
\cos^{2}{\varphi_{2}}
\\
&
C_{(In)}^{A,H}(\mathbf{r}_{1}; \mathbf{r}_{2}) 
=
\frac{A_{1,2}}{2}
\sin^{2}{\varphi_{1}}\cos^{2}{\varphi_{2}}^{2}
\end{aligned}
\right\rbrace \quad
\mathcal{V}_{AH}(\varphi_{1}; \varphi_{2})
= 
\cos{(2 \varphi_{1})}
\end{equation}

\begin{equation}
\left.
\begin{aligned}
&
C_{(Out)}^{A,V}(\mathbf{r}_{1}; \mathbf{r}_{2})
=
\frac{A_{1,2}}{4}
\sin^{2}{\varphi_{2}}
\\
&
C_{(In)}^{A,V}(\mathbf{r}_{1}; \mathbf{r}_{2}) 
=
\frac{A_{1,2}}{2}
\cos^{2}{\varphi_{1}}\sin^{2}{\varphi_{2}}
\end{aligned}
\right\rbrace \quad
\mathcal{V}_{AV}(\varphi_{1}; \varphi_{2}) = 
- \cos{(2 \varphi_{1})}
\end{equation}
In case of A-A, A-D polarisation configurations, the visibility map depends on the $\varphi_{1}$ and the $\varphi_{2}$ angles, both.
\begin{equation}
\left.
\begin{aligned}
&
C_{(Out)}^{A,A}(\mathbf{r}_{1}; \mathbf{r}_{2})
=
\frac{A_{1,2}}{8}
\left[
\cos^{2}{(\varphi_{1}-\varphi_{2})} + \sin^{2}{(\varphi_{1}+\varphi_{2})}
\right]
\\
&
C_{(In)}^{A,A}(\mathbf{r}_{1}; \mathbf{r}_{2})
=
\frac{A_{1,2}}{4}
\sin^{2}{(\varphi_{1}+\varphi_{2})}
\end{aligned}
\right\rbrace \quad
\mathcal{V}_{AA}(\varphi_{1}; \varphi_{2}) = 
\frac{
\cos^{2}{(\varphi_{1}-\varphi_{2})} - \sin^{2}{(\varphi_{1}+\varphi_{2})}
}{
\cos^{2}{(\varphi_{1}-\varphi_{2})} + \sin^{2}{(\varphi_{1}+\varphi_{2})}
}
\end{equation}
\begin{equation}
\left.
\begin{aligned}
&
C_{(Out)}^{A,D}(\mathbf{r}_{1}; \mathbf{r}_{2})
=
\frac{A_{1,2}}{8}
\left[
\cos^{2}{(\varphi_{1}+\varphi_{2})} + \sin^{2}{(\varphi_{1}-\varphi_{2})}
\right]
\\
&
C_{(In)}^{A,D}(\mathbf{r}_{1}; \mathbf{r}_{2}) 
=
\frac{A_{1,2}}{4}
\sin^{2}{(\varphi_{1}-\varphi_{2})}
\end{aligned}
\right\rbrace \quad
\mathcal{V}_{AD}(\varphi_{1}; \varphi_{2}) = \frac{
\cos^{2}{(\varphi_{1}+\varphi_{2})} - \sin^{2}{(\varphi_{1}-\varphi_{2})}
}{
\cos^{2}{(\varphi_{1}+\varphi_{2})} + \sin^{2}{(\varphi_{1}-\varphi_{2})}
}
\end{equation}
In order to obtain the coincidences that what we effectively see on the camera when a photon is recorded on the bucketdetector, we have to integrate these expressions over the full $\varphi_{2}$ azimuthal angle, thus obtaining $C^{\alpha, \beta}_{(In)}(\mathbf{d}_{1,2}; \varphi_{1})$ and $C^{\alpha, \beta}_{(Out)}(\mathbf{d}_{1,2}; \varphi_1)$, with $\mathbf{d}_{1,2} = (r_{1}, z_{1}; r_{2}, z_{2})$.
From these expressions we calculate the associated visibility distribution 
\begin{equation}
\mathcal{V}(\varphi_{1}) = 
\frac{
C^{\alpha, \beta}_{(Out)}(\mathbf{d}_{1,2}; \varphi_{1}) - C^{\alpha, \beta}_{(In)}(\mathbf{d}_{1,2}; \varphi_{1}) 
}{
C^{\alpha, \beta}_{(Out)}(\mathbf{d}_{1,2}; \varphi_{1})
}
\end{equation}
The spatial visibilities obtained for the 8 different polarisers configurations are
\begin{equation}
\begin{aligned}
&
\mathcal{V}_{HH} = 1
\\
&
\mathcal{V}_{HV} = -1
\\
&
\mathcal{V}_{AH} = \cos{(2\varphi_1)}
\\
&
\mathcal{V}_{AV} = - \cos{(2\varphi_1)} \\
&
\mathcal{V}_{HA} = \mathcal{V}_{HD} = \mathcal{V}_{AA} = \mathcal{V}_{AD} = 0,
\end{aligned}
\label{Eq:StructuredVis}
\end{equation}
and coincide with the ones listd in eq. (6) in the main text.
The visibility spatial distributions for $AH$ and $AV$ are shown in Fig. \ref{fig: Visibility Maps}.

\bibliographystyle{naturemag}
\bibliography{biblio.bib}